\documentclass{PoS}

\usepackage{amsmath,amssymb,bm}

\title{Higher dimensional HQET parameters}

\ShortTitle{Higher dimensional HQET parameters}

\author{\speaker{Gil Paz}\\
Department of Physics and Astronomy \\
Wayne State University\\
Detroit, Michigan 48201, USA \\
                      E-mail: \email{gilpaz@wayne.edu}}

\abstract{Effective field  theories such as Heavy Quark Effective Theory (HQET) are indispensable tools in controlling the effects of the strong interaction. The increasing experimental precision requires the knowledge of higher dimensional operators.  We present a general method that allows for an easy construction of HQET  operators that contain two heavy quarks and any number of covariant derivatives. 
}

\FullConference{ 9th International Workshop on the CKM Unitarity Triangle\\
                 28  November - 3 December 2016\\
                 Tata Institute for Fundamental Research (TIFR), Mumbai, India}

\begin{document}

\section{Motivation}
Inclusive semileptonic $B$ decays and the $Q_{7\gamma}-Q_{7\gamma}$ contribution to $B\to X_s\gamma$ can be described by a local operator product expansion: $\Gamma=\sum_{n=0}^\infty m_b^{-n}\sum_k c_{k,n}\langle O_{k,n}\rangle
$ see the talk ``Theory of radiative $B$ decays" in these proceedings. The Wilson  coefficients $c_{k,n}$ are perturbative. The matrix elements $\langle O_{k,n}\rangle$ of Heavy Quark Effective Theory (HQET) operators are non-perturbative and often called HQET parameters. Higher dimensional HQET parameters are of phenomenological interest. For example, $|V_{cb}|$ extraction from inclusive $B$ decays \cite{Gambino:2016jkc} uses dimension 7 and 8 HQET operators \cite{Mannel:2010wj}. 

Several questions arise. 1) Are these \emph{all} the possible HQET operators at dimension 7 and 8? or do inclusive $B$ decays require a subset of the possible operators? 2) Can we construct higher dimensional HQET operators? 3) Since HQET and Non Relativistic QCD (NRQCD) are related \cite{Manohar:1997qy, Gunawardana:2017zix}, what are the corresponding  NRQCD operators? 4) What can we learn  about the structure of effective field theories (EFTs)? We will answer all of these questions during this talk.

\section{A little bit of history}
Because of  the relation between HQET, NRQCD, and Non Relativistic QED (NRQED), the question of possible operators at each dimension can be traced to the early days of quantum mechanics. The time and space components of the covariant derivative are $D_t=\partial/\partial t+ieA^0,\,\bm{D}=\bm\nabla-ie\bm A$. Schr\"odinger equation contain the operator $i D_t  + \bm{D}^2/{2 M}$. When discussing hydrogen fine structure one encounters operators such as spin-orbit coupling: $ \bm{\sigma}\cdot \bm{B}$, relativistic correction: $\bm{D}^4$, and the Darwin term: $\bm{\partial}\cdot \bm{E}$, where $\bm{E}=(-i/g)[D^0,\bm D]$ and $\bm{B}^i=\epsilon^{ijk}(i/2g)[\bm{D}^j,\bm{D}^k]$. Organizing the operators in a Lagrangian form and keeping operators up to dimension 6, one obtains the dimension-6 NRQED Lagrangian first presented in 1986 \cite{Caswell:1985ui}:
\begin{equation}\label{NRQED6}
{\cal L}_{\mbox{\scriptsize NRQED}}^{\mbox{\scriptsize dim$\leq$ 6}} = \psi^\dagger
  \bigg\{   i D_t  + {\bm{D}^2 \over 2 M} 
  c_F g{ \bm{\sigma}\cdot \bm{B} \over 2M}   
+ c_D g{ [\bm{\partial}\cdot \bm{E} ]  \over 8 M^2}  
+ i c_S g{ \bm{\sigma}
    \cdot ( \bm{D} \times \bm{E} - \bm{E}\times \bm{D} ) \over 8 M^2} 
   \bigg\} \psi.
 \end{equation}

In 1994 the first systematic discussion of HQET parameters was presented in \cite{Mannel:1994kv}. Between HQET fields $\bar h\dots h$ the Dirac basis reduces to  $\{1, \bm\sigma \}=\{1,s^\lambda \}$ with $v\cdot s=0$, where $v$ is the velocity. The most general bilinear HQET operator is of the form $\bar h\, iD^{\mu_1}\dots\, iD^{\mu_n}(s^\lambda)h$. This operator vanishes if it is contracted with $v_{\mu_1}$, $v_{\mu_n}$, or $v_{\lambda}$ \cite{Mannel:1994kv}. Consider matrix elements of such operators between heavy meson pseudo-scalar states $H$. One finds one dimension 3 operator with no derivatives: $\bar hh$ and dimension 4 operators with one derivatives: $\bar h\, iD^{\mu}(s^\lambda)h$ that have vanishing matrix elements. For dimension 5 and 6 one finds \cite{Mannel:1994kv}  
\begin{eqnarray}\label{HQET56}
\langle H|
\bar h (iD_\alpha) (iD_\beta) h
| H\rangle  &=& 2 M_H [g_{\alpha \beta} - v_\alpha v_\beta ]
\frac{1}{3}\lambda_1 \nonumber\\
\langle H|
\bar h (iD_\alpha) (iD_\beta) s_\lambda h
| H\rangle
&=&   2 M_H d_H i \varepsilon_{\nu \alpha \beta \lambda } v^\nu
\frac{1}{6}\lambda_2\nonumber\\
\langle H|
\bar h (iD_\alpha) (iD_\mu)  (iD_\beta) h_v
| H\rangle   
&=&  2 M_H [g_{\alpha \beta} - v_\alpha v_\beta ] v_\mu
\frac{1}{3}\rho_1 \nonumber\\
\langle H|
\bar h (iD_\alpha) (iD_\mu) (iD_\beta) s_\lambda h
| H\rangle 
&=& 2M_H d_H i \varepsilon_{\nu \alpha \beta \lambda } v^\nu v_\mu
\frac{1}{6} \rho_2 .
\end{eqnarray}
The same source also discussed higher dimensional operators, but unfortunately the enumeration of the operators is incorrect.  

For both dimension 5 and 6 we have two HQET parameters corresponding to two operators: a Spin-Independent (SI) operator and a Spin-Dependent (SD) operator. Notice that this is the same number of operator as in (\ref{NRQED6}). This is not an accident. While HQET and NRQCD differ in their kinetic terms: ${\cal L}^{kinetic}_{HQET}=\bar hiv\cdot D\, h, \,
{\cal L}^{kinetic}_{NRQCD}=\psi^\dagger \left( i D_t  + {\bm{D}^2/2 M} \right)\psi$, and power counting, there is a correspondence between HQET and NRQCD (NRQED) operators \cite{Manohar:1997qy,Gunawardana:2017zix}. For example:  \\ 
\begin{table}[h]
\begin{center}
\begin{tabular}{ccc}
& NRQED (1920's-1980's)&HQET (1990's)\\\\
Dimension 5&${ \bm{D}^2}$ &$(iD_\perp)^2$\\ 
&$ { \bm{\sigma}\cdot \bm{B}}$ &$(iD_\perp^\mu)( iD_\perp^\nu)(-i\sigma^{\mu\nu})$\\ 
Dimension 6&$ {  [\bm{\partial}\cdot \bm{E} ]}$ &$(iD_{\perp\mu}) (iv\cdot D)( iD_\perp^\mu)$\\ 
&$ {   \bm{\sigma}
    \cdot ( \bm{D} \times \bm{E} - \bm{E}\times \bm{D} )}$ &$(iD_\perp^\mu)( iv\cdot D)(iD_\perp^\nu)(-i\sigma^{\mu\nu})$\\ 
\end{tabular} 
\end{center}
\caption{\label{correspondence} Correspondence between dimension 5 and 6 HQET and NRQCD (NRQED) operators. The notation is  $\sigma^{\mu\nu}=i[\gamma^\mu,\gamma^\nu]/2$ and $D_\perp^\mu=D^\mu-v\cdot Dv^\mu$.}
\end{table}
  
In 1997 the NRQCD and HQET Lagrangian up to dimension 7 was given in \cite{Manohar:1997qy}. The Lagrangian contains six SI and five SD operators. Two pairs of the SI operators have the same Lorentz structure but different color structures. Two of the SD operators containing $\bm\sigma\cdot(\bm{B}\times\bm{B})$ and $\bm\sigma\cdot(\bm{E}\times\bm{E})$ vanish for NRQED. 

The dimension 7 contribution to semileptonic decays was discussed in 2006 in \cite{Dassinger:2006md}  and again in 2010 in \cite{Mannel:2010wj}. How many operators do we have at dimension 7? For SI operators \cite{Mannel:1994kv} lists two, \cite{Manohar:1997qy} lists six,  \cite{Dassinger:2006md}  lists three,  and \cite{Mannel:2010wj} lists four. For SD operators \cite{Mannel:1994kv} lists five, \cite{Manohar:1997qy} lists five,  \cite{Dassinger:2006md}  lists two,  and \cite{Mannel:2010wj} lists five.  HQET-NRQCD correspondence implies six SI and five SD operators.  Why were only four of them needed in \cite{Mannel:2010wj}?

For dimension 8 operators \cite{Mannel:2010wj} lists seven SI operators and eleven SD operators. In 2012 the dimension 8 NRQED Lagrangian was given in \cite{Hill:2012rh}.  It lists four SI operators and eight SD operators. Comparing to  \cite{Mannel:2010wj} the difference is persumeably NRQCD operators that vanish for NRQED.

In all of the papers \cite{Caswell:1985ui, Mannel:1994kv, Manohar:1997qy, Dassinger:2006md, Mannel:2010wj, Hill:2012rh} there is no derivation of the operators for each dimension. We learn from this history that finding all of the HQET and NRQCD operators at a given dimension is not easy. Is there a systematic way to do that?

\section{Higher dimensional HQET parameters}
As shown in \cite{Gunawardana:2017zix} the answer is yes . We consider matrix elements of the form  $\langle H|\bar h\, iD^{\mu_1}\dots\, iD^{\mu_n}(s^\lambda)h|H\rangle$ and decompose them in terms of the tensors  $v^{\mu_i}$, $g^{\mu_i\mu_j}$, and $\epsilon^{\alpha\beta\rho\sigma}$, subject to constraints from Parity and Time reversal symmetry ($PT$), Hermitian conjugation, and the fact that we are working in four dimensions. Finally, we also need to check for possible multiple color structures. 

Parity and time reversal are symmetries of HQET.  In particular under their combined operation we have $p=(p^0,\vec p\,)\stackrel{PT}{\to}(p^0,\vec p\,)=p\Rightarrow v=p/m\stackrel{PT}{\to} v$, $iD^\mu\stackrel{PT}{\to} iD^\mu$,  $\bar h  h\stackrel{PT}{\to} \bar h h$, and $\bar h s^\lambda h\stackrel{PT}{\to} -\,\bar h s^\lambda h$. Since $T$ is anti-linear we have in total 
\begin{eqnarray}
\langle H|\bar h\, iD^{\mu_1}\dots\, iD^{\mu_n}h|H\rangle\stackrel{PT}{=}
\langle H|\bar h\, iD^{\mu_1}\dots\, iD^{\mu_n}h|H\rangle^*\nonumber\\
\langle H|\bar h\, iD^{\mu_1}\dots\, iD^{\mu_n}s^\lambda h|H\rangle\stackrel{PT}{=}
-\langle H|\bar h\, iD^{\mu_1}\dots\, iD^{\mu_n}s^\lambda h|H\rangle^*\,.
\end{eqnarray}
Therefore matrix elements of SI (SD) operators are real (imaginary).

Since $\bar h h$, $\bar h s^\lambda h$, $iD^\mu$ are hermitian, using Hermitian conjugation  we find that 
\begin{equation}
\langle H|\bar h\, iD^{\mu_1}\dots\, iD^{\mu_n}(s^\lambda)h|H\rangle=\langle H|\left(\bar h\, iD^{\mu_1}\dots\, iD^{\mu_n}(s^\lambda)h\right)^\dagger|H\rangle^*=\langle H|\bar h\, iD^{\mu_n}\dots\, iD^{\mu_1}(s^\lambda)h|H\rangle^*\,.
\end{equation}
Combining this with the $PT$ constraints we find that under the inversion of the indices matrix elements of SI operators are symmetric and matrix elements of SD operators are anti-symmetric.

$H$ is a pseudo-scalar so its matrix elements can only depend on the tensors  $v^{\mu_i}, g^{\mu_i\mu_j},$ and $\epsilon^{\alpha\beta\rho\sigma}$.  Alternatively, following \cite{Mannel:2010wj} we define  $\Pi^{\mu\nu}=g^{\mu\nu}-v^\mu v^\nu$. For the standard choice of $v=(1,0,0,0)$: $\Pi^{00}=0$ and $\Pi^{ij}=-\delta^{ij}$. Since the indices in $\epsilon^{\alpha\beta\rho\sigma}$ cannot all be orthogonal to $v$, we can replace  $\epsilon^{\alpha\beta\rho\sigma}$ by $\epsilon^{\alpha\beta\rho\sigma}v_\alpha$. In the following we use the tensors  $v^{\mu_i}, \Pi^{\mu_i\mu_j},$ and $\epsilon^{\alpha\beta\rho\sigma}v_\alpha$. 

Another constraint arises from the fact that we are working in four dimensions. As a result not all tensors with more than four indices are independent. For example,  for the dimension 7 SD HQET operators we need the tensor $\Pi^{\mu\nu}\epsilon^{\alpha\beta\rho\sigma}v_\alpha$.  Three of its indices are the same and tensors obtained by permuting its indices are not linearly independent. Analogous constraint for SI operators start at dimension 11 where we need $\Pi^{\mu_1\mu_2}\Pi^{\mu_3\mu_4}\Pi^{\mu_5\mu_6}\Pi^{\mu_7\mu_8}$. Since all of the indices are space-like, we must have four identical indices. 

Finally, as was pointed out by Kobach and Pal in \cite{Kobach:2017xkw}, starting at dimension 7 there can be multiple color structures for operators with the same Lorentz structure. This occurs when we have an anti-commutator of pure color-octet operators. Using the color identity $\left\{T^a,T^b\right\}=\frac13\delta^{ab}+ d^{abc}T^c$ we find two possible color structures. It is more convenient to use  $\left\{T^a,T^b\right\}$ and $\delta^{ab}$ as a basis instead of $\delta^{ab}$ and $d^{abc}$. Operators with $\left\{T^a,T^b\right\}$ color structure are generated by commutator and anti-commutators of covariant derivatives and appear at tree level when analyzing power corrections for inclusive $B$ decays. As explained in \cite{Gunawardana:2017zix}, operators with $\delta^{ab}$ color structure arise only at one-loop level and are beyond the current needed level of precision for $B$ physics applications.

Using these general considerations one can list the various HQET parameters that appear in the decomposition of the general HQET operator of a given dimension. The dimension 5 and 6 decompositions are listed in (\ref{HQET56}). Let us find the  decomposition of the dimension 7 SI operator. The matrix element is $\langle H |\bar h\, iD^{\mu_1}iD^{\mu_2}iD^{\mu_3}iD^{\mu_4}h|H\rangle$. Consider its decomposition in terms possible tensors. We can have a product of two $\Pi$'s or a product of $\Pi$ and two $v$'s. For products of two $\Pi$'s  we can contract $\mu_1$ with $\mu_2,\mu_3,$ or $\mu_4$ using $\Pi$. The other two indices are also contracted  by $\Pi$. In total we have three such combinations of two $\Pi$'s. Using two $v$'s, they can only be contracted  with $\mu_2$ and $\mu_3$ giving us a fourth tensor. In total we have 
\begin{eqnarray}\label{SI7}
\dfrac1{2M_H}\langle H |\bar h\, iD^{\mu_1}iD^{\mu_2}iD^{\mu_3}iD^{\mu_4}h|H\rangle&=&a_{12}^{(7)}\Pi^{\mu_1\mu_2}\Pi^{\mu_3\mu_4}+a_{13}^{(7)}\Pi^{\mu_1\mu_3}\Pi^{\mu_2\mu_4}+\nonumber\\
&+&a_{14}^{(7)}\Pi^{\mu_1\mu_4}\Pi^{\mu_2\mu_3}+b^{(7)}\Pi^{\mu_1\mu_4}v^{\mu_2} v^{\mu_3}. 
\end{eqnarray}   

Possible multiple color structures arise from operators of the from $\bar h\left\{[iD^{\mu_i},iD^{\mu_j}],[iD^{\mu_k},iD^{\mu_l}]\right\}h$. We can form scalar operators by multiplying these structures by one of the four possible tensors on the right hand side of (\ref{SI7}). We find  only two linearly independent combinations from all of the contractions, implying that two of the operators have two possible color structures each. In total we have six linearly independent possible SI operators, but only four are needed at tree level. This explains the difference between \cite{Mannel:2010wj} and \cite{Manohar:1997qy}. 

Similarly one can find the decomposition of the SD operator:
\begin{eqnarray}\label{SD7}
&&\dfrac1{2M_H}\langle H |\bar h\, iD^{\mu_1}iD^{\mu_2}iD^{\mu_3}iD^{\mu_4}s^\lambda h|H\rangle=i\tilde a_{12}^{(7)}\left(\Pi^{\mu_1\mu_2}\epsilon^{\rho\mu_3\mu_4\lambda}v_{\rho}-\Pi^{\mu_4\mu_3}\epsilon^{\rho\mu_2\mu_1\lambda}v_{\rho}\right)+\nonumber\\
&&+i\tilde a_{13}^{(7)}\left(\Pi^{\mu_1\mu_3}\epsilon^{\rho\mu_2\mu_4\lambda}v_{\rho}-\Pi^{\mu_4\mu_2}\epsilon^{\rho\mu_3\mu_1\lambda}v_{\rho}\right)+i\tilde a_{14}^{(7)}\Pi^{\mu_1\mu_4}\epsilon^{\rho\mu_2\mu_3\lambda}v_{\rho}+\nonumber\\
&+&i\tilde a_{23}^{(7)}\Pi^{\mu_2\mu_3}\epsilon^{\rho\mu_1\mu_4\lambda}v_{\rho}+i\tilde b^{(7)}v^{\mu_2}v^{\mu_3}\epsilon^{\rho\mu_1\mu_4\lambda}v_{\rho}.
\end{eqnarray}
Checking possible multiple color structures as for the SI operators, we find that there are  none. This explains the agreement about the number of SD dimension 7 operators between \cite{Manohar:1997qy}  and \cite{Mannel:2010wj}.

The dimension 8 decomposition follows along similar lines, see \cite{Gunawardana:2017zix} for details. We find that there are eight SI operators in total, but two of them only differ in their color structure. This implies that only seven of them are needed at tree level, in agreement with  \cite{Mannel:2010wj}.  Similarly, there are seventeen SD operators and six of them only differ in their color structure. This implies that only eleven of them are needed at tree level, in agreement with  \cite{Mannel:2010wj}. We can also go beyond known results in the literature.  For example, for the dimension 9 HQET operators 
\cite{Gunawardana:2017zix}: 
\begin{eqnarray} \label{dim9}
&&\dfrac1{2M_H}\langle H |\bar h\, iD^{\mu_1}iD^{\mu_2}iD^{\mu_3}iD^{\mu_4}iD^{\mu_5}iD^{\mu_6}h|H\rangle=a^{(9)}_{12,34}\,\Pi^{\mu_1\mu_2}\Pi^{\mu_3\mu_4}\Pi^{\mu_5\mu_6}+\nonumber\\
&&+ a^{(9)}_{12,35}\,
\left(\Pi^{\mu_1\mu_2}\Pi^{\mu_3\mu_5}\Pi^{\mu_4\mu_6}+\Pi^{\mu_1\mu_3}\Pi^{\mu_2\mu_4}\Pi^{\mu_5\mu_6}\right)+a^{(9)}_{12,36}\,\left(\Pi^{\mu_1\mu_2}\Pi^{\mu_3\mu_6}\Pi^{\mu_4\mu_5}+\Pi^{\mu_1\mu_4}\Pi^{\mu_2\mu_3}\Pi^{\mu_5\mu_6}\right)+\nonumber\\
&&+ a^{(9)}_{13,25}\,\Pi^{\mu_1\mu_3}\Pi^{\mu_2\mu_5}\Pi^{\mu_4\mu_6}+a^{(9)}_{13,26}\,\left(\Pi^{\mu_1\mu_3}\Pi^{\mu_2\mu_6}\Pi^{\mu_4\mu_5}+\Pi^{\mu_1\mu_5}\Pi^{\mu_2\mu_3}\Pi^{\mu_4\mu_6}\right)+a^{(9)}_{14,25}\,\Pi^{\mu_1\mu_4}\Pi^{\mu_2\mu_5}\Pi^{\mu_3\mu_6}+\nonumber\\
&&+a^{(9)}_{14,26}\,
\left(\Pi^{\mu_1\mu_4}\Pi^{\mu_2\mu_6}\Pi^{\mu_3\mu_5}+\Pi^{\mu_1\mu_5}\Pi^{\mu_2\mu_4}\Pi^{\mu_3\mu_6}\right)+a^{(9)}_{15,26}\,\Pi^{\mu_1\mu_5}\Pi^{\mu_2\mu_6}\Pi^{\mu_3\mu_4}+a^{(9)}_{16,23}\,\Pi^{\mu_1\mu_6}\Pi^{\mu_2\mu_3}\Pi^{\mu_4\mu_5}\nonumber\\
&&+a^{(9)}_{16,24}\,\Pi^{\mu_1\mu_6}\Pi^{\mu_2\mu_4}\Pi^{\mu_3\mu_5}
+a^{(9)}_{16,25}\,\Pi^{\mu_1\mu_6}\Pi^{\mu_2\mu_5}\Pi^{\mu_3\mu_4}+b^{(9)}_{12,36}\,\left(\Pi^{\mu_1\mu_2}\Pi^{\mu_3\mu_6}v^{\mu_4}v^{\mu_5}+\Pi^{\mu_1\mu_4}\Pi^{\mu_5\mu_6}v^{\mu_2}v^{\mu_3}\right)+\nonumber\\
&&+b^{(9)}_{12,46}\,\left(\Pi^{\mu_1\mu_2}\Pi^{\mu_4\mu_6}v^{\mu_3}v^{\mu_5}+\Pi^{\mu_1\mu_3}\Pi^{\mu_5\mu_6}v^{\mu_2}v^{\mu_4}\right)+b^{(9)}_{12,56}\,\Pi^{\mu_1\mu_2}\Pi^{\mu_5\mu_6}v^{\mu_3}v^{\mu_4}+\nonumber\\
&&+b^{(9)}_{13,26}\,\left(\Pi^{\mu_1\mu_3}\Pi^{\mu_2\mu_6}v^{\mu_4}v^{\mu_5}+\Pi^{\mu_1\mu_5}\Pi^{\mu_4\mu_6}v^{\mu_2}v^{\mu_3}\right)+b^{(9)}_{13,46}\,\Pi^{\mu_1\mu_3}\Pi^{\mu_4\mu_6}v^{\mu_2}v^{\mu_5}+\nonumber\\
&&+b^{(9)}_{14,26}\,\left(\Pi^{\mu_1\mu_4}\Pi^{\mu_2\mu_6}v^{\mu_3}v^{\mu_5}+\Pi^{\mu_1\mu_5}\Pi^{\mu_3\mu_6}v^{\mu_2}v^{\mu_4}\right)+b^{(9)}_{14,36}\,\Pi^{\mu_1\mu_4}\Pi^{\mu_3\mu_6}v^{\mu_2}v^{\mu_5}+b^{(9)}_{15,26}\,\Pi^{\mu_1\mu_5}\Pi^{\mu_2\mu_6}v^{\mu_3}v^{\mu_4}+\nonumber\\
&&b^{(9)}_{16,23}\,\left(\Pi^{\mu_1\mu_6}\Pi^{\mu_2\mu_3}v^{\mu_4}v^{\mu_5}+\Pi^{\mu_1\mu_6}\Pi^{\mu_4\mu_5}v^{\mu_2}v^{\mu_3}\right) +b^{(9)}_{16,24}\,\left(\Pi^{\mu_1\mu_6}\Pi^{\mu_2\mu_4}v^{\mu_3}v^{\mu_5}+\Pi^{\mu_1\mu_6}\Pi^{\mu_3\mu_5}v^{\mu_2}v^{\mu_4}\right)+\nonumber\\
&&+b^{(9)}_{16,25}\,\Pi^{\mu_1\mu_6}\Pi^{\mu_2\mu_5}v^{\mu_3}v^{\mu_4}+b^{(9)}_{16,34}\,\Pi^{\mu_1\mu_6}\Pi^{\mu_3\mu_4}v^{\mu_2}v^{\mu_5}+c^{(9)}\,\Pi^{\mu_1\mu_6}v^{\mu_2}v^{\mu_3}v^{\mu_4}v^{\mu_5}.
\end{eqnarray}
In principle there can be multiple color structures too, but at the current needed level of precision only operators with $\left\{T^a,T^b\right\}$ color structure are needed. 

An interesting question is what are the Wilson coefficients of the operators. In particular, the relations between coefficients of operators of different dimensions. These are known as ``reparameterization invariance"  \cite{Luke:1992cs}, or more accurately, ``Lorentz invariance" constraints \cite{Heinonen:2012km}. For NRQED they are known up to dimension 8 operators \cite{Heinonen:2012km, Hill:2012rh}, but not for NRQCD or HQET.  Such relations allow to determine the contribution of some higher dimensional operators based on lower dimensional ones. This has applications to semileptonic and radiative $B$ decays, see e.g. \cite{Becher:2007tk, Ewerth:2009yr, Manohar:2010sf}.

\section{Conclusions} 
We presented a general method to construct HQET operators by using the tensor decomposition of HQET matrix elements. There are several applications to this method. First, the tensor decomposition allows to easily relate different bases\footnote{The appendix of  \cite{Heinonen:2016cwm} also lists a tensor decomposition for HQET operators up to dimension 8 and relates it to \cite{Mannel:2010wj}.}. In \cite{Gunawardana:2017zix} we relate it to the dimension 7 operators basis of \cite{Manohar:1997qy} and \cite{Mannel:2010wj} and the dimension 8  operators basis of \cite{Mannel:2010wj}. Second, we present for the first time the decomposition of the general SI dimension 9 HQET operator, equation (\ref{dim9}). Third, in \cite{Gunawardana:2017zix} moments of the leading power shape function are calculated up to and including dimension 9 HQET operators. This can improve the parameterization of the shape function relevant to the extraction of $|V_{ub}|$. Fourth,  in \cite{Gunawardana:2017zix} we present the full dimension 8 NRQCD Lagrangian. 

We can now answer the questions posed earlier. 1) Are the dimension 7 and 8 HQET operators in \cite{Mannel:2010wj} all the possible operators? No, there are other operators that appear with ${\cal O}(\alpha_s)$  Wilson coefficients, beyond the current needed level of precision. 2) Can we construct higher dimensional HQET operators? Yes, we gave the general method for that. 3) What are the corresponding  NRQCD operators? The answer is in \cite{Gunawardana:2017zix}. 4) What can we learn about the structure of EFTs? They are simpler than we might think. This is also the conclusion of \cite{Henning:2015alf} for the Standard Model EFT.

 \section*{Acknowledgments} 
 This work was supported by DOE grant DE-SC0007983.

\end{document}